\documentclass[aps,twocolumn]{revtex4}%
\usepackage{amsfonts}
\usepackage{amsmath}
\usepackage{amssymb}
\usepackage{graphicx}%
\begin{document}
\preprint{ }
\title{Radiation pressure driven vibrational modes in ultra-high-Q silica microspheres}
\author{R. Ma, A. Schliesser, P. Del'Haye, A. Dabirian, T. J. Kippenberg}
\affiliation{Max-Planck-Institut f\"{u}r Quantenoptik, Hans-Kopfermann-Str. 1, 85748
Garching, Germany}

\begin{abstract}
Quantitative measurements of the vibrational eigenmodes in ultra-high-Q silica
microspheres are reported. The modes are efficiently excited via
radiation-pressure induced dynamical back-action of light confined in the
optical whispering-gallery modes of the microspheres (i.e. via the parametric
oscillation instability). Two families of modes are studied and their
frequency dependence on sphere size investigated. The measured frequencies are
in good agreement both with Lamb's theory and numerical finite element
simulation and are found to be proportional to the sphere's inverse diameter.

\end{abstract}
\maketitle

Silica microcavities\cite{Vahala2003} such as
microspheres\cite{Braginskii1987} or microtoroids\cite{Armani2003} possess
ultra-high-Q optical whispering gallery modes (WGMs), while simultaneously
exhibiting mechanical modes which lie typically in the radio frequency range.
Owing to the resonant buildup of light within these cavities the effect of
radiation pressure is enhanced, leading to mutual coupling between the
mechanical and optical modes, as first predicted by Braginsky in the context
of the Laser Interferometer Gravitational Wave Observatory,
LIGO\cite{Braginskii2001}. When entering the regime where the photon lifetime
is comparable to the mechanical oscillation period and the cavity is pumped
with a laser whose frequency slightly exceeds the WGM resonance (i.e.
blue-detuned excitation), this mutual coupling gives rise to a parametric
oscillation instability\cite{Braginskii2001} which is characterized by
regenerative mechanical oscillation of the mechanical eigenmodes. This
phenomenon has been first reported in toroid microcavities
\cite{Rokhsari2005,Kippenberg2005,Carmon2005}. On the other hand, red-detuned
light can induce cooling of the mechanical mode, as recently
reported\cite{Gigan2006,Arcizet2006,Schliesser2006}. In this letter,
parametric oscillation instability in ultra-high-Q silica microspheres is
observed and the mechanical resonant frequencies and mode patterns studied. In
contrast to earlier studies of acoustic modes of
nanospheres\cite{Lim2004,Kuok2003} using Raman or Brillouin scattering from
ensembles, the present method allows measurement of the mechanical modes of
single microspheres in a larger diameter regime (35-110 $\mu m$ in our case).
Furthermore, the mechanical Q-factors are determined.
\begin{figure}[ptb]
\centerline{\includegraphics[width=7.6cm]{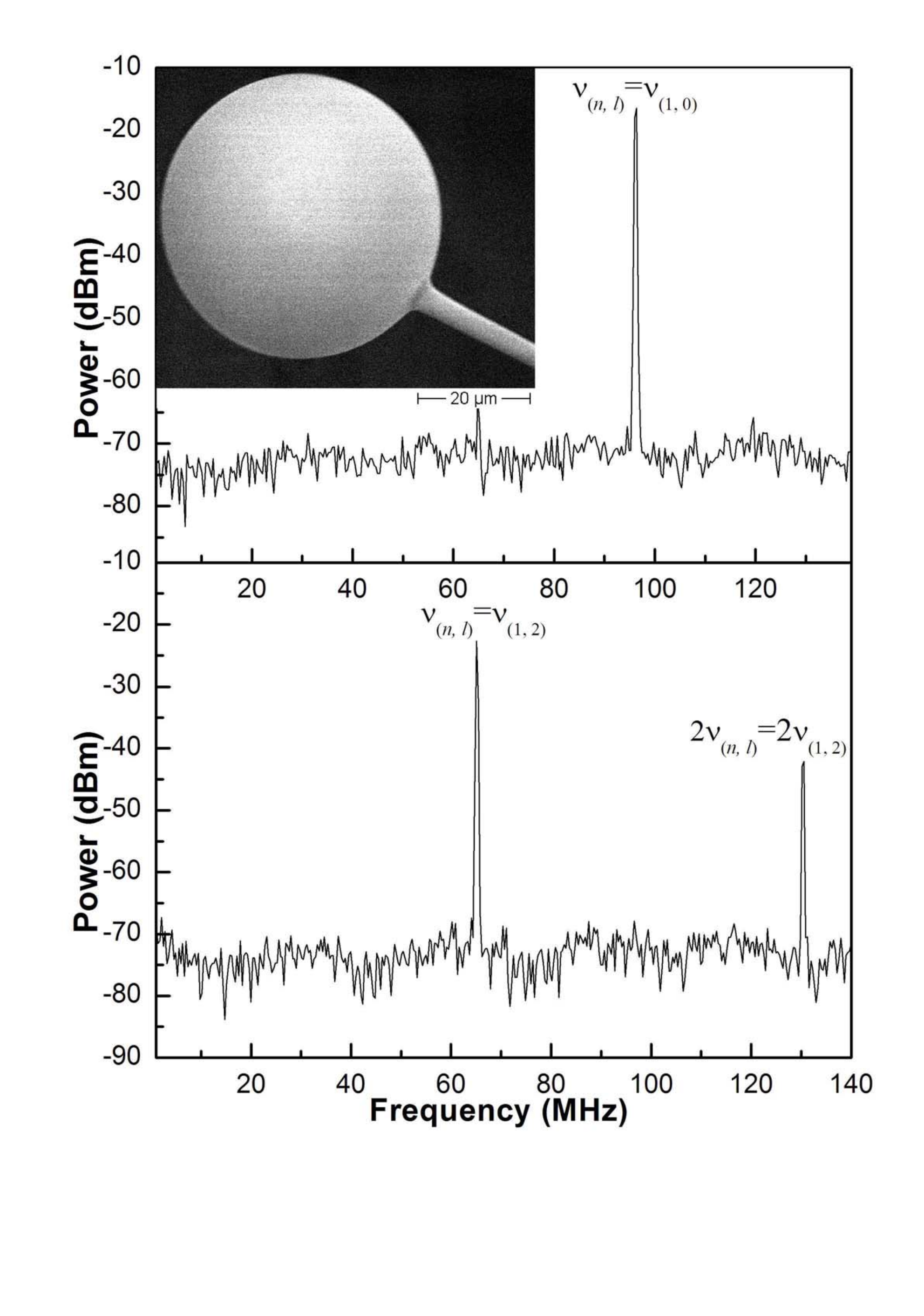}}\caption{Spectrum
of the transmitted optical laser of the fiber taper when coupled to a silica
microsphere with a diameter of 49 $\mu m$. Two families of spheroidal
mechanical modes could be driven regeneratively. Specifically, the modes are
indentified as $\nu_{(1,2)}$ (a quadrupole mode) and $\nu_{(1,0)}$(a radial
breathing mode). The launched power in this experiment was 600 micro-Watts and
was sufficient to exceed the threshold for parametric oscillation instability.
The inset shows the SEM image of the microsphere.}%
\end{figure}

Ultra high-Q ($Q>10^{8}$) silica microspheres are fabricated by melting the
tip of a single mode optical fiber with a $CO_{2}$ laser ($\lambda=10.6$ $\mu
$m). Due to the strong absorption of silica around 10.6 $\mu m$, surface
tension induces highly symmetric silica spheres with a near-atomically smooth
surface\cite{Braginskii1990}. The sphere is held by a thin fiber stem (cf.
Fig. 1 inset). The WGM are excited with high ideality\cite{Spillane2002} by
evanescent coupling via a tapered optical fiber \cite{Cai2000} using a
$1550-$nm tunable external cavity diode laser as pump source. Owing to the
high finesse (approximately $10^{5}$), the large optical energy stored in the
microcavity exerts a force on the cavity sidewalls due to radiation pressure.
This force can give rise to regeneratively driven mechanical oscillations if
photon lifetime is similar to the inverse acoustic resonance
frequency\cite{Kippenberg2005,Rokhsari2005}. In essence, the radial force
exerted by radiation pressure takes the cavity out of resonance by deformation
of the cavity wall which causes subsequently a reduction in the radiation
pressure force. The whole process resumes upon the restoration of the original
shape of the cavity, leading to a periodic motion of the cavity. In this way,
modes with radial deformation can be excited which modify the path length of
the optical wave and therefore affect the magnitude of the radiation force. In
the present experiments the threshold for the parametric instability was in
the range of typically 100 micro-Watts. The driven mechanical oscillations
causes the appearance of motional sidebands (and their harmonics) which can be
readily detected in the spectrum of the transmitted laser light using an
electronic spectrum analyzer. Fig. 1. shows a typical spectrum of the
transmission, showing regenerative oscillations of two different mechanical
eigen-modes of the same sphere under different taper loading conditions. By
changing the taper loading (and hence the optical Q) the value of the
oscillation threshold of the two modes crosses, thereby causes a switching
from the low- to a high frequency mode as coupling strength increases and the
optical linewidth decreases, in agreement with the theoretical
predictions\cite{Rokhsari2006}. To identify the mode families, the vibrational
frequencies of the two lowest lying mechanical frequencies were recorded as a
function of size. While observable in principle, light-induced modifications
of the mechanical modes' dynamical properties\cite{Schliesser2006}, in
particular shifts in the mechanical resonance frequency, are considered to be
small (relative frequency shift typically $<0.1\%$) and are neglected in this
letter. The result of this study is shown in Fig. 2. As evident, the spheres
mechanical frequencies are inversely proportional to the sphere
diameter.\begin{figure}[ptb]
\centerline{\includegraphics[width=8.6cm]{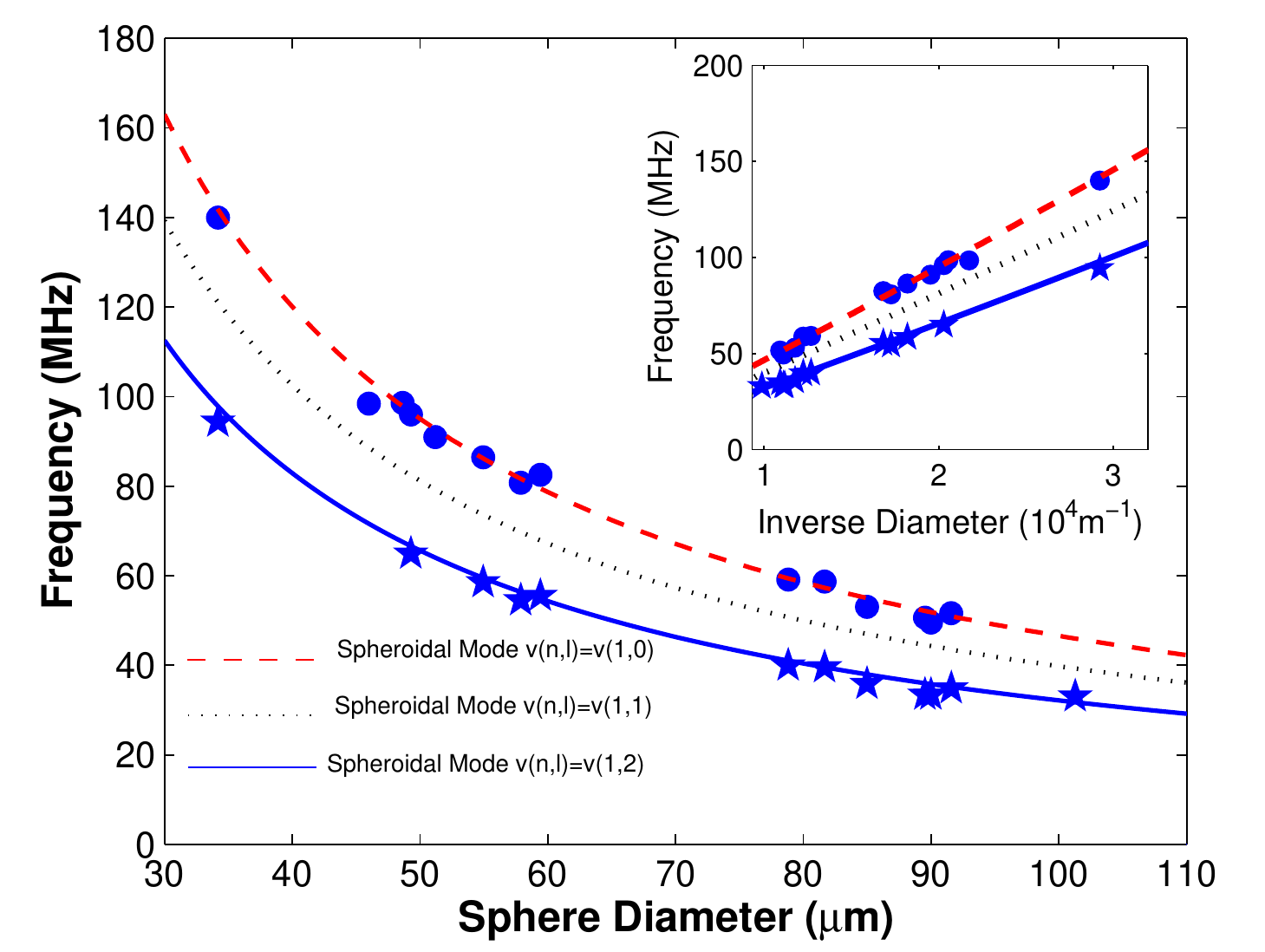}}\caption{Experimentally
measured frequencies of the spheroidal modes, with stars denoting the
$\nu_{(1,2)}$ mode and dots denoting the $\nu_{(1,0)}$ mode. Numerically
calculated eigenfrequencies of these modes are shown in blue solid line
($\nu_{(1,2)}$) and red dashed line ($\nu_{(1,0)}$). Spheroidal mode
$\nu_{(1,1)}$ with the frequencies lie between $\nu_{(1,2)}$ and $\nu_{(1,0)}$
, which is not experimentally observed, is also presented in black dotted
line. The inset shows the relationship between the frequencies of the
eigenmodes and the inverse diameter of the silica microspheres.}%
\end{figure}

Next, the observed modes where identified by numerical studies. Since the
fabricated silica microspheres exhibited no observable eccentricity under SEM
imaging (see the inset of Fig. 1), and the diameter ratio of the stem holding
the sphere and the sphere itself is on the order of 0.1, the sphere can be
considered as almost free; hence it is a judicious choice to adopt the
stress-free boundary condition. Studies on the nature of the fundamental modes
of vibration for small elastic spheres with free-surface boundary condition
are well known. The first well-established theory was formulated by Lamb, with
two types of modes predicted, the spheroidal and torsional
modes\cite{Lamb1884}. The equation describing the wave propagation in a
homogeneous elastic body with free surface can be written
as\cite{Nishiguchi1981},
\begin{equation}
\rho\ddot{\mathbf{u}}=(\lambda+2\mu)\nabla(\nabla\cdot\mathbf{u})-\mu
\nabla\times(\nabla\times\mathbf{u})
\end{equation}

where \textbf{u} is the displacement vector, $\rho$ is the mass density, and
$\lambda$ and $\mu$ are $Lam\acute{e}$ constants. Here $\lambda\equiv
\frac{\sigma E}{(1+\sigma)(1-2\sigma)}$ , and $\mu\equiv\frac{E}{2(1+\sigma)}$
, with $E$ denoting the Young's modulus and $\sigma$ the Poisson ratio of the
material. Eq. (1) can be solved by introducing a scalar potential $\phi_{0}$
and two vector potentials $\Phi_{1}=(r\phi_{1},0,0)$ and $\Phi_{2}=(r\phi
_{2},0,0)$ with $\mathbf{u}=\nabla\phi_{0}+\nabla\times\Phi_{1}+\nabla
\times\nabla\times\Phi_{2}$. Then the general solutions of the equations
resulting from (1) are written as
\begin{equation}
\phi_{2}=\sum_{\ell,m}A_{i}^{(\ell,m)}j_{\ell}(\frac{2\pi\nu_{n,\ell
,m}\mathbf{r}}{V_{I}})Y_{\ell}^{m}(\theta,\Psi)e^{-2\pi i\nu_{n,\ell,m}t}%
\end{equation}
where $j_{l}$ is the spherical Bessel function and $Y_{\ell}^{m}$ is the
spherical harmonic function and $V_{0}$ is the longitudinal sound velocity and
$V_{1}=V_{2}$ are the transverse sound velocities. An angular momentum mode
number $\ell$ ( $\ell=0,1,2..$), an azimuthal mode number m ($-\ell\leq
m\leq\ell$) and a radial mode number $n$ $(n=1,2,..)$ are used to characterize
the acoustic modes, where $n=1$ corresponds to the surface mode and $n\geq2$
to inner modes and $\nu_{n,\ell,m}\ $denotes the frequency of the vibration
characterized by the mode numbers $(n,\ell,m)$. It is noteworthy that a
spheroidal mode with angular momentum $\ell$, is $(2\ell+1)$-fold degenerate,
hence in Fig. 1 we use $(n,\ell)$ instead of $(n,\ell,m)$ to assign the eigenfrequencies.

Two classes of modes are derived when applying the free boundary
condition\cite{Nishiguchi1981}. One of them is the torsional vibration which
induces only shear stress without volume change, and no radial displacement
takes place in these modes. Thus these modes cannot be excited using radiation
pressure which relies on a change in\ the optical path length. In contrast,
the class of mode in which volume change is present is referred to as
spheroidal modes. According to Lambs theory the $\ell=0$ spheroidal mode
eigenvalue equation is written as
\begin{equation}
\frac{tan(hR)}{hR}-\frac{1}{1-\frac{1}{4}(k^{2}/h^{2})h^{2}R^{2}}=0
\end{equation}
where $k=2\pi\nu/V_{1},$ $h=2\pi\nu/V_{0}$, $R$ is the radius of the sphere,
$V_{0}=\sqrt{(\lambda+2\mu)/\rho}$, and $V_{1}=V_{2}=\sqrt{\mu/\rho}$. Other
eigenvalue equations for torsional modes and $\ell>0$ spheroidal modes are
contained in Ref.\cite{Lamb1884}. Next, the resonant frequencies of the lowest
frequencies (i.e. $n=1$, $\ell=0,1,2$) were numerically calculated as a
function of sphere size (compare Fig. 2, solid lines). The eigenvalues versus
the inverse of the diameters are shown in the inset. As seen from Fig. 2 the
measured data fit very well to the theoretical prediction based on the
$\nu_{(1,0)}$ (radial breathing) and $\nu_{(1,2)}$ (quadrupole) mode, and
reveals that the eigen-frequencies of the microspheres have linear dependence
on the inverse microsphere diameter.

We note that the sphere's eccentricity can lift the $(2\ell+1)$-fold
degeneracy in the mode number $m$ of the $\nu_{(1,2)}$ mode\cite{Tamura1982},
which in the experiments was not observed owing to the high degree of symmetry
of the microspheres.\begin{figure}[ptb]
\centerline{\includegraphics[width=7.6cm] {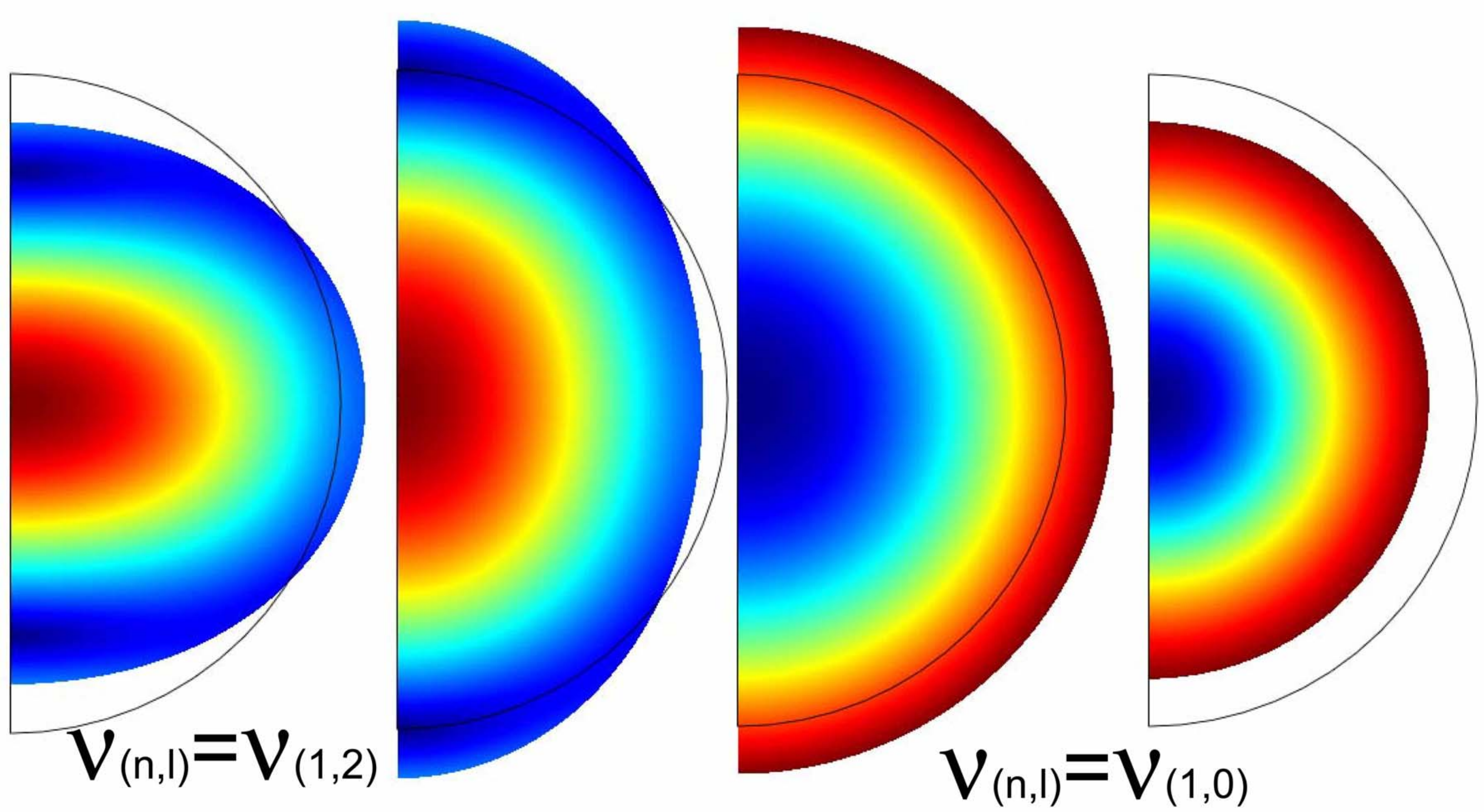}}\caption{Finite
element modeling of three spheroidal modes $\nu_{(1,2)}$ (left) and
$\nu_{(1,0)}$(right) of a silica microsphere with its von Mises stress (color
coded) and deformed shape (greatly exaggerated for clarity). }%
\end{figure}

It is worth pointing out that $\ell=1$ spheroidal mode has the eigenfrequency
lying between $\ell=0$ and $\ell=2$ spheroidal modes (Fig. 2); however, it is
not observed in the experiment since $\ell=1$ spheroidal mode result in a
vanishing path length change around the optical whispering gallery mode
trajectory such that the mutual coupling of optical and mechanical mode vanishes.

To gain a more complete understanding of the mechanical modes, the numerical
studies were complemented with finite element simulations of the stress and
strain fields. Using axial symmetric finite element modeling, the mode
families $\nu_{(1,0)}$ and $\nu_{(1,2)}$ were calculated and excellent
agreement found with the numerical solution of the preceding section. In the
simulation, a Young's modulus of $73.1\times10^{9}$ Pa, a Poisson ratio of
0.17 and a density of the microsphere of $2.203\times10^{3}$ $kg/m^{3}$ is
used (Corresponding to $\lambda=16.09\times10^{9}$ Pa and $\mu=31.24\times
10^{9}$ Pa). The longitudinal and transverse sound velocities are $5972$ m/s
and $3755$ m/s, respectively. Fig. 3 depicts the deformed shape and von Mises
stress, as obtained from finite element simulations of two spheroidal modes
with the lowest frequencies. The stresses are greatly exaggerated for clarity.

To determine the mechanical dissipation of the vibrational modes, the
mechanical quality factor of the sphere modes where measured in the
sub-threshold regime (in a purged nitrogen environment). For this measurement
the laser power was adjusted to a level far below the threshold of the
mechanical oscillation. In addition, the mechanical Q factor was measured for
both blue-detuning (i.e. where the radiation pressure decreases the mechanical
dissipation, causing mechanical amplification) and red-detuning (where the
radiation pressure causes the mechanical damping to be increased, leading to
cooling). Both values coincided closely, and yielded Q-values of up to 7000
for the radial breathing mode. Note that the Q was difficult to measure for
the $\ell=2$ mode owing to the 5-fold degeneracy of the mode which precludes
fitting of the spectrum with a single oscillator function.

In conclusion, spheroidal acoustic modes in silica microspheres driven by
radiation pressure induced parametric oscillation instability are reported.
The observed vibrational modes are indentified as spheroidal modes whose
frequencies agree well with Lamb's theory and numerical simulation and exhibit
a linear dependence on the inverse of the sphere diameters.

\subsection{Acknowledgements}

We thank Prof. Dr. Grundler and Dr. Berberich in the physics department of
Technical University of Munich for providing access to the SEM. This work was
funded via a Max Planck Independent Junior Research Group grant and a Marie
Curie International Reintegration, a Marie Curie Excellence Grant (RG-UHQ) and
the NIM Initiative.


\newpage\bigskip
\begin{thebibliography}{19}
\expandafter\ifx\csname natexlab\endcsname\relax\def\natexlab#1{#1}\fi
\expandafter\ifx\csname bibnamefont\endcsname\relax
  \def\bibnamefont#1{#1}\fi
\expandafter\ifx\csname bibfnamefont\endcsname\relax
  \def\bibfnamefont#1{#1}\fi
\expandafter\ifx\csname citenamefont\endcsname\relax
  \def\citenamefont#1{#1}\fi
\expandafter\ifx\csname url\endcsname\relax
  \def\url#1{\texttt{#1}}\fi
\expandafter\ifx\csname urlprefix\endcsname\relax\def\urlprefix{URL }\fi
\providecommand{\bibinfo}[2]{#2}
\providecommand{\eprint}[2][]{\url{#2}}

\bibitem[{\citenamefont{Vahala}(2003)}]{Vahala2003}
\bibinfo{author}{\bibfnamefont{K.~J.} \bibnamefont{Vahala}},
  \bibinfo{journal}{Nature} \textbf{\bibinfo{volume}{424}},
  \bibinfo{pages}{839} (\bibinfo{year}{2003}).

\bibitem[{\citenamefont{Braginskii and Ilchenko}(1987)}]{Braginskii1987}
\bibinfo{author}{\bibfnamefont{V.~B.} \bibnamefont{Braginskii}}
  \bibnamefont{and} \bibinfo{author}{\bibfnamefont{V.~S.}
  \bibnamefont{Ilchenko}}, \bibinfo{journal}{Doklady Akademii Nauk Sssr}
  \textbf{\bibinfo{volume}{293}}, \bibinfo{pages}{1358} (\bibinfo{year}{1987}).

\bibitem[{\citenamefont{Armani et~al.}(2003)\citenamefont{Armani, Kippenberg,
  Spillane, and Vahala}}]{Armani2003}
\bibinfo{author}{\bibfnamefont{D.~K.} \bibnamefont{Armani}},
  \bibinfo{author}{\bibfnamefont{T.~J.} \bibnamefont{Kippenberg}},
  \bibinfo{author}{\bibfnamefont{S.~M.} \bibnamefont{Spillane}},
  \bibnamefont{and} \bibinfo{author}{\bibfnamefont{K.~J.}
  \bibnamefont{Vahala}}, \bibinfo{journal}{Nature}
  \textbf{\bibinfo{volume}{421}}, \bibinfo{pages}{925} (\bibinfo{year}{2003}).

\bibitem[{\citenamefont{Braginsky et~al.}(2001)\citenamefont{Braginsky,
  Strigin, and Vyatchanin}}]{Braginskii2001}
\bibinfo{author}{\bibfnamefont{V.~B.} \bibnamefont{Braginsky}},
  \bibinfo{author}{\bibfnamefont{S.~E.} \bibnamefont{Strigin}},
  \bibnamefont{and} \bibinfo{author}{\bibfnamefont{S.~P.}
  \bibnamefont{Vyatchanin}}, \bibinfo{journal}{Physics Letters A}
  \textbf{\bibinfo{volume}{287}}, \bibinfo{pages}{331} (\bibinfo{year}{2001}).

\bibitem[{\citenamefont{Rokhsari et~al.}(2005)\citenamefont{Rokhsari,
  Kippenberg, Carmon, and Vahala}}]{Rokhsari2005}
\bibinfo{author}{\bibfnamefont{H.}~\bibnamefont{Rokhsari}},
  \bibinfo{author}{\bibfnamefont{T.~J.} \bibnamefont{Kippenberg}},
  \bibinfo{author}{\bibfnamefont{T.}~\bibnamefont{Carmon}}, \bibnamefont{and}
  \bibinfo{author}{\bibfnamefont{K.~J.} \bibnamefont{Vahala}},
  \bibinfo{journal}{Optics Express} \textbf{\bibinfo{volume}{13}},
  \bibinfo{pages}{5293} (\bibinfo{year}{2005}).

\bibitem[{\citenamefont{Kippenberg et~al.}(2005)\citenamefont{Kippenberg,
  Rokhsari, Carmon, Scherer, and Vahala}}]{Kippenberg2005}
\bibinfo{author}{\bibfnamefont{T.~J.} \bibnamefont{Kippenberg}},
  \bibinfo{author}{\bibfnamefont{H.}~\bibnamefont{Rokhsari}},
  \bibinfo{author}{\bibfnamefont{T.}~\bibnamefont{Carmon}},
  \bibinfo{author}{\bibfnamefont{A.}~\bibnamefont{Scherer}}, \bibnamefont{and}
  \bibinfo{author}{\bibfnamefont{K.~J.} \bibnamefont{Vahala}},
  \bibinfo{journal}{Physical Review Letters} \textbf{\bibinfo{volume}{95}},
  \bibinfo{pages}{033901} (\bibinfo{year}{2005}).

\bibitem[{\citenamefont{Carmon et~al.}(2005)\citenamefont{Carmon, Rokhsari,
  Yang, Kippenberg, and Vahala}}]{Carmon2005}
\bibinfo{author}{\bibfnamefont{T.}~\bibnamefont{Carmon}},
  \bibinfo{author}{\bibfnamefont{H.}~\bibnamefont{Rokhsari}},
  \bibinfo{author}{\bibfnamefont{L.}~\bibnamefont{Yang}},
  \bibinfo{author}{\bibfnamefont{T.~J.} \bibnamefont{Kippenberg}},
  \bibnamefont{and} \bibinfo{author}{\bibfnamefont{K.~J.}
  \bibnamefont{Vahala}}, \bibinfo{journal}{Physical Review Letters}
  \textbf{\bibinfo{volume}{94}} (\bibinfo{year}{2005}).

\bibitem[{\citenamefont{Gigan et~al.}(2006)\citenamefont{Gigan, Bohm,
  Paternostro, Blaser, Langer, Hertzberg, Schwab, Bauerle, Aspelmeyer, and
  Zeilinger}}]{Gigan2006}
\bibinfo{author}{\bibfnamefont{S.}~\bibnamefont{Gigan}},
  \bibinfo{author}{\bibfnamefont{H.~R.} \bibnamefont{Bohm}},
  \bibinfo{author}{\bibfnamefont{M.}~\bibnamefont{Paternostro}},
  \bibinfo{author}{\bibfnamefont{F.}~\bibnamefont{Blaser}},
  \bibinfo{author}{\bibfnamefont{G.}~\bibnamefont{Langer}},
  \bibinfo{author}{\bibfnamefont{J.~B.} \bibnamefont{Hertzberg}},
  \bibinfo{author}{\bibfnamefont{K.~C.} \bibnamefont{Schwab}},
  \bibinfo{author}{\bibfnamefont{D.}~\bibnamefont{Bauerle}},
  \bibinfo{author}{\bibfnamefont{M.}~\bibnamefont{Aspelmeyer}},
  \bibnamefont{and}
  \bibinfo{author}{\bibfnamefont{A.}~\bibnamefont{Zeilinger}},
  \bibinfo{journal}{Nature} \textbf{\bibinfo{volume}{444}}, \bibinfo{pages}{67}
  (\bibinfo{year}{2006}).

\bibitem[{\citenamefont{Arcizet et~al.}(2006)\citenamefont{Arcizet, Cohadon,
  Briant, Pinard, and Heidmann}}]{Arcizet2006}
\bibinfo{author}{\bibfnamefont{O.}~\bibnamefont{Arcizet}},
  \bibinfo{author}{\bibfnamefont{P.~F.} \bibnamefont{Cohadon}},
  \bibinfo{author}{\bibfnamefont{T.}~\bibnamefont{Briant}},
  \bibinfo{author}{\bibfnamefont{M.}~\bibnamefont{Pinard}}, \bibnamefont{and}
  \bibinfo{author}{\bibfnamefont{A.}~\bibnamefont{Heidmann}},
  \bibinfo{journal}{Nature} \textbf{\bibinfo{volume}{444}}, \bibinfo{pages}{71}
  (\bibinfo{year}{2006}).

\bibitem[{\citenamefont{Schliesser et~al.}(2006)\citenamefont{Schliesser,
  Del'Haye, Nooshi, Vahala, and Kippenberg}}]{Schliesser2006}
\bibinfo{author}{\bibfnamefont{A.}~\bibnamefont{Schliesser}},
  \bibinfo{author}{\bibfnamefont{P.}~\bibnamefont{Del'Haye}},
  \bibinfo{author}{\bibfnamefont{N.}~\bibnamefont{Nooshi}},
  \bibinfo{author}{\bibfnamefont{K.~J.} \bibnamefont{Vahala}},
  \bibnamefont{and} \bibinfo{author}{\bibfnamefont{T.~J.}
  \bibnamefont{Kippenberg}}, \bibinfo{journal}{Physical Review Letters}
  \textbf{\bibinfo{volume}{97}} (\bibinfo{year}{2006}).

\bibitem[{\citenamefont{Lim et~al.}(2004)\citenamefont{Lim, Kuok, Ng, and
  Wang}}]{Lim2004}
\bibinfo{author}{\bibfnamefont{H.~S.} \bibnamefont{Lim}},
  \bibinfo{author}{\bibfnamefont{M.~H.} \bibnamefont{Kuok}},
  \bibinfo{author}{\bibfnamefont{S.~C.} \bibnamefont{Ng}}, \bibnamefont{and}
  \bibinfo{author}{\bibfnamefont{Z.~K.} \bibnamefont{Wang}},
  \bibinfo{journal}{Applied Physics Letters} \textbf{\bibinfo{volume}{84}},
  \bibinfo{pages}{4182} (\bibinfo{year}{2004}).

\bibitem[{\citenamefont{Kuok et~al.}(2003)\citenamefont{Kuok, Lim, Ng, Liu, and
  Wang}}]{Kuok2003}
\bibinfo{author}{\bibfnamefont{M.~H.} \bibnamefont{Kuok}},
  \bibinfo{author}{\bibfnamefont{H.~S.} \bibnamefont{Lim}},
  \bibinfo{author}{\bibfnamefont{S.~C.} \bibnamefont{Ng}},
  \bibinfo{author}{\bibfnamefont{N.~N.} \bibnamefont{Liu}}, \bibnamefont{and}
  \bibinfo{author}{\bibfnamefont{Z.~K.} \bibnamefont{Wang}},
  \bibinfo{journal}{Physical Review Letters} \textbf{\bibinfo{volume}{90}}
  (\bibinfo{year}{2003}).

\bibitem[{\citenamefont{Braginskii et~al.}(1990)\citenamefont{Braginskii,
  Ilchenko, and Gorodetskii}}]{Braginskii1990}
\bibinfo{author}{\bibfnamefont{V.~B.} \bibnamefont{Braginskii}},
  \bibinfo{author}{\bibfnamefont{V.~S.} \bibnamefont{Ilchenko}},
  \bibnamefont{and} \bibinfo{author}{\bibfnamefont{M.~L.}
  \bibnamefont{Gorodetskii}}, \bibinfo{journal}{Uspekhi Fizicheskikh Nauk}
  \textbf{\bibinfo{volume}{160}}, \bibinfo{pages}{157} (\bibinfo{year}{1990}).

\bibitem[{\citenamefont{Spillane et~al.}(2003)\citenamefont{Spillane,
  Kippenberg, Painter, and Vahala}}]{Spillane2002}
\bibinfo{author}{\bibfnamefont{S.~M.} \bibnamefont{Spillane}},
  \bibinfo{author}{\bibfnamefont{T.~J.} \bibnamefont{Kippenberg}},
  \bibinfo{author}{\bibfnamefont{O.~J.} \bibnamefont{Painter}},
  \bibnamefont{and} \bibinfo{author}{\bibfnamefont{K.~J.}
  \bibnamefont{Vahala}}, \bibinfo{journal}{Physical Review Letters}
  \textbf{\bibinfo{volume}{91}}, \bibinfo{pages}{art. no.}
  (\bibinfo{year}{2003}).

\bibitem[{\citenamefont{Cai et~al.}(2000)\citenamefont{Cai, Painter, and
  Vahala}}]{Cai2000}
\bibinfo{author}{\bibfnamefont{M.}~\bibnamefont{Cai}},
  \bibinfo{author}{\bibfnamefont{O.}~\bibnamefont{Painter}}, \bibnamefont{and}
  \bibinfo{author}{\bibfnamefont{K.~J.} \bibnamefont{Vahala}},
  \bibinfo{journal}{Physical Review Letters} \textbf{\bibinfo{volume}{85}},
  \bibinfo{pages}{74} (\bibinfo{year}{2000}).

\bibitem[{\citenamefont{Rokhsari et~al.}(2006)\citenamefont{Rokhsari,
  Kippenberg, Carmon, and Vahala}}]{Rokhsari2006}
\bibinfo{author}{\bibfnamefont{H.}~\bibnamefont{Rokhsari}},
  \bibinfo{author}{\bibfnamefont{I.~J.} \bibnamefont{Kippenberg}},
  \bibinfo{author}{\bibfnamefont{T.}~\bibnamefont{Carmon}}, \bibnamefont{and}
  \bibinfo{author}{\bibfnamefont{K.~J.} \bibnamefont{Vahala}},
  \bibinfo{journal}{Ieee Journal of Selected Topics in Quantum Electronics}
  \textbf{\bibinfo{volume}{12}}, \bibinfo{pages}{96} (\bibinfo{year}{2006}).

\bibitem[{\citenamefont{Lamb}(1884)}]{Lamb1884}
\bibinfo{author}{\bibfnamefont{H.}~\bibnamefont{Lamb}}, \bibinfo{journal}{Proc.
  London. Math. Soc.} p. \bibinfo{pages}{189} (\bibinfo{year}{1884}).

\bibitem[{\citenamefont{Nishiguchi and Sakuma}(1981)}]{Nishiguchi1981}
\bibinfo{author}{\bibfnamefont{N.}~\bibnamefont{Nishiguchi}} \bibnamefont{and}
  \bibinfo{author}{\bibfnamefont{T.}~\bibnamefont{Sakuma}},
  \bibinfo{journal}{Solid State Communications} \textbf{\bibinfo{volume}{38}},
  \bibinfo{pages}{1073} (\bibinfo{year}{1981}).

\bibitem[{\citenamefont{Tamura et~al.}(1982)\citenamefont{Tamura, Higeta, and
  Ichinokawa}}]{Tamura1982}
\bibinfo{author}{\bibfnamefont{A.}~\bibnamefont{Tamura}},
  \bibinfo{author}{\bibfnamefont{K.}~\bibnamefont{Higeta}}, \bibnamefont{and}
  \bibinfo{author}{\bibfnamefont{T.}~\bibnamefont{Ichinokawa}},
  \bibinfo{journal}{Journal of Physics C-Solid State Physics}
  \textbf{\bibinfo{volume}{15}}, \bibinfo{pages}{4975} (\bibinfo{year}{1982}).

\end{thebibliography}
\end{document}